# Single-Photon-Level Quantum Image Memory Based on Cold Atomic Ensembles

Dong-Sheng Ding, Zhi-Yuan Zhou, Bao-Sen Shi,[*] and Guang-Can Guo

*Key Laboratory of Quantum Information, University of Science and Technology of China, Hefei 230026, China*

*Corresponding author:* [*]*drshi@ustc.edu.cn*

**An important step towards the successful development of network that allows the distribution of quantum information is the storage of light in a matter at the single-photon level. Encoding photons in high-dimensional photonic states can significantly increase their information-carrying capacity. Realizing a quantum memory that can record the shape of a single photon is a challenging work. So far, there is no any report on the storage of an image at true single photon level. In this work, we report on the first experimental realization of storing a true single photon carrying an orbital angular momentum via electromagnetically induced transparency in a cold atomic ensemble and retrieved later after a given storage time. We experimentally not only prove the preservation of the property of the single photon, but also show very good similarity on the spatial structures of the input and the retrieved photons. More importantly, we also demonstrate the preservation of the coherence of the single photon during the storage using a Sagnac interferometer. The demonstrated capability of storing a spatial structure at single-photon-level opens the possibility to the realization of a high-dimensional quantum memory.**

A light with an orbital angular momentum (OAM) has many exciting applications, including optical communications [1,2], trapping of particles [3,4] and astrophysics [5,6]. In quantum information and quantum optics fields, a light with an OAM has been used to encode the information in a high-dimensional state [7,8], by which the network's information-carrying capacity can be significantly increased. Besides, higher-dimensional states enable more efficient quantum-information processing, and large-alphabet quantum key distribution affords an more secure flux of information [9], etc. Establishing a quantum network involves the coherent interaction [10] between a light and a matter. There are some experiments based on such a light-matter interface, for example, establishing the entanglement of OAM between a photon and a collective atomic spin excitation [11,12] and storing a light in a matter [13-21]. Recently, storing a light carrying an OAM or a spatial structure via electromagnetically induced transparency (EIT) in an atomic ensemble [13-16] or in a cryogenically cooled doped crystal [17], or via gradient echo technique in an atomic ensemble [18-20], or using atomic frequency comb technique in solids doped with rare-earth-metal ions [21] has been realized, but these important works involve bright lights. Very recently, there were some works related to the storage of a light carrying an OAM or a spatial structure near single-photon-level [22,23], but the light is a strongly attenuated laser, not a true single photon. So far, there is no any report on the storage of an image at true single photon level. The reversible transfer of a quantum state between a true single photon and a matter is an essential requirement in quantum information science. It is a crucial resource for the implementation of quantum repeater, a potential solution to overcome the limited distance of quantum communication schemes due to losses in optical fibers. Ref. 24 theoretically shows that the transverse degrees of freedom combined with the longitudinal ones constitut a valuable resource for multimode quantum memories with excellent capacities. Here, we report on the first experimental realization of a multimode optical memory at the true single-photon-level via EIT in a cold atomic ensemble. In our experiment, we prepare a non-classical correlated photon pair using the spontaneous four-wave-mixing (SFWM) via a double lambda configuration in a cold $Rb^{85}$ atomic ensemble. One photon of each pair is used as a trigger and the other is stored in the experiment. The photon to be stored carries an OAM 1 per photon in units of $\hbar$, is mapped into and stored in another cold atomic ensemble via EIT firstly, then is retrieved after a programmed storage time. We experimentally not only prove that the non-classical correlation between the trigger photon and the retrieved photon is kept, but also demonstrate that the spatial structure of the photon is also preserved very well during this storage process. More important, we prove the preservation of the coherence of the single photon during the storage process by the aid of a Sagnac interferometer. Combining these results with the recent important progresses made in quantum repeater, we could expect the possible establishment of a high-dimensional quantum network in the future.

The single photon used in the experiment was prepared by using SFWM in a cold $^{85}$Rb atomic ensemble trapped in a two-dimensional magneto-optical traps (MOT) [25]. The simplified experimental setup was shown in Fig. 1(a). Signal 1 photon at 780 nm was used to be a trigger and signal 2 photon at

795 nm was stored in the follow experiments, therefore we called the signal 1 photon the trigger and the signal 2 photon signal here. We firstly proved the existence of the non-classical correlation between these two photons by demonstrating the strong violation of the Cauchy-Schwarz inequality [26]. Besides, we also proved the single photon property of the signal photon by performing the Hanbury Brown-Twiss experiment with a trigger photon [27]. Experimentally we obtained the value α of about 0.025±0.005 for the signal photon. For an ideally prepared single photon state, anti-correlation parameter α→0 [27, 28]. The detail about the experiment was shown in method section.

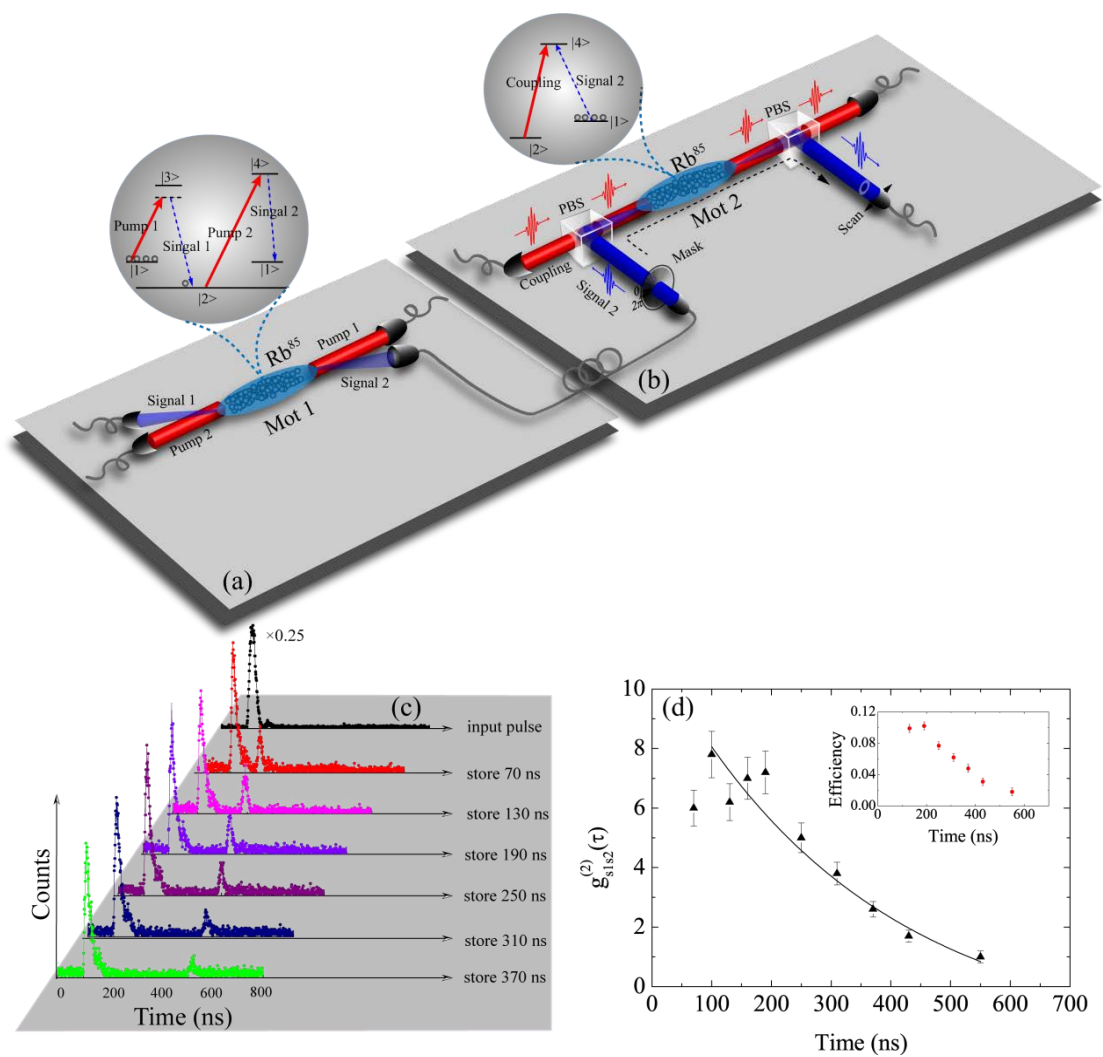

Fig. 1 (Color online.) (a) the simplified diagram of generating a non-classical correlation photon by using SFWM. (b) the storage diagram. PBS: Glan-Taylor polarization beam splitter with the extinction ratio of $10^5$:1. (c) the experimental results of the coincidence between the retrieved signal and the trigger against the storage time. (d) the cross-correlation function $g_{s1,s2}(\tau)$ between the retrieved signal and the trigger photons vs. the storage time. The solid line was fitted curve by an exponential decay

function of $g^{(2)}_{s1s2}(\tau)=Ae^{-\tau/T}+g_0$ (where A=13.3, T=348, $g_0$=-1.89). The inset showed the efficiency function against the storage time.

We firstly performed the experimental storage of a single photon not carrying a spatial structure through EIT in the second cold $^{85}$Rb atomic ensemble, the simplified setup was shown in Fig. 1(b). The cross-coincidence counts between the trigger photon and the leaked signal photon and the retrieved signal photon were measured respectively (Fig. 1(c)). We also measured the efficiency of storage against the storage time, the results were shown in Fig. 1(d). The inset showed the efficiency function against the storage time, the maximum efficiency obtained was ~10%. We could find that even after about 400 ns storage, there still existed the strong non-classical correlation between the retrieved signal photon and the trigger photon. The detail about the experiment was shown in method section too. Furthermore, we checked the single photon property of the signal photon after the storage by performing the HanburyBrown-Twiss experiment with a trigger photon again [27]. Experimentally we obtained the value α of about 0.32±0.08 for the retrieved signal photon after it had been stored for about 190 ns, which clearly confirmed that the single photon property was preserved during the storage process.

In order to store a photon carrying a spatial structure, we inserted a spiral phase plate (VPP-1c, RPC Photonics, transmission coefficient >95%) into the optical route along which the signal photon transmitted, thus signal photon carried a donut-shaped spatial structure shown in Fig. 2(a) ( the signal had a well-defined OAM of 1 in units of $\hbar$ ). After had been stored for about 100 ns in MOT2, the signal photon was retrieved and collected into a single mode fiber, the tip of the fiber was scanned along the transverse direction. Before performing the storage experiment, we made the coupling efficiency of the fiber in different point along the transverse direction balanced. During the experiment, we firstly scanned the transverse position of the input signal photon and measured the cross-correlation function between the input signal and the trigger photons, obtained a donut-shaped curve shown in Fig. 2 (b), where Fig. 2(a) was the image of a laser beam passing through the spiral phase plate directly, taken by a common CCD camera. The power distribution along transverse direction was shown on the top of Fig. 2(a). The Fig. 2(c) was the cross-correlation function between the retrieved signal and the trigger photons, which also had a donut-shaped curve. In order to compare Fig. 2(b) with Fig. 2(c), we made two calculations: visibility and similarity. The visibility of the image can be calculated using the formula $V=\frac{g_{s1,s2,max}-g_{s1,s2,min}}{g_{s1,s2,max}+g_{s1,s2,min}}$, where $g_{s1,s2,max}$ ($g_{s1,s2,min}$) was the maximal (minimal) cross-correlation value in Fig. 2. According to the data shown in 2(b) and 2(c) of Fig. 2, we calculated the visibility, which was of 0.9 for input signal and of 0.88 for the retrieved signal respectively. We also analyzed the fidelity of the retrieved image by calculating the similarity R of the retrieved image compared with the inputimage using the formula $R=\frac{\sum_m\sum_n A_{mn}B_{mn}}{\sqrt{(\sum_m\sum_n A_{mn}^2)(\sum_m\sum_n B_{mn}^2)}}$, where A and B were the gray-scale matrixes of the two images to be compared [20]. The high similarity meant the high fidelity. The calculated similarity of the retrieved image was 0.996. (In our calculation, m equaled to a constant because we only scanned the tip of the fiber along transverse direction.).

The figures presented in Fig.2 clearly provided the experimental evidence that an image memory at the true single-photon level can be realized using a cold atomic ensemble, the main features of the image had been preserved during the storage process. Besides, the non-classical correlation between the trigger photon and the retrieved photon was also kept. This point was crucial for establishing a high-dimensional quantum repeater. In this experiment, the main noise which reduced the signal-to-noise ratio was from the photon generated through the atomic transition of $5P_{1/2}(F'=3) \rightarrow 5S_{1/2}(F=3)$. In order to get rid of it, we had to make the atomic population in state |2> be near zero before performing storage experiment, this was achieved by the follow way: at the end of the trapping cycle and before coupling laser was turn on, the atoms were pumped to the state |1> by turning off the trapping laser 0.5 ms ahead the repump laser. The dephasing between two ground states induced by the earth's magnetic field had effect on the storage process, this was clearly shown in Fig. 1(d), where the non-classical correlation between the photons and the retrieved signal intensity decreased with the increment of the storage time. The storage efficiency could be increased by controlling both single-photon wave packet and the memory bandwidth [29, 30]. One main point which affected the quality of the retrieved image was the atomic diffusion. It would soften the edge of an image [22, 31]. A possible way to solute this problem was using a 4-f imaging system in the experiment, by this way, the Fourier transform of the image, instead of the image itself, was stored in the atomic ensemble. By this configuration, diffusion can be reduced significantly [22, 31] and the image could be stored for a more long time.

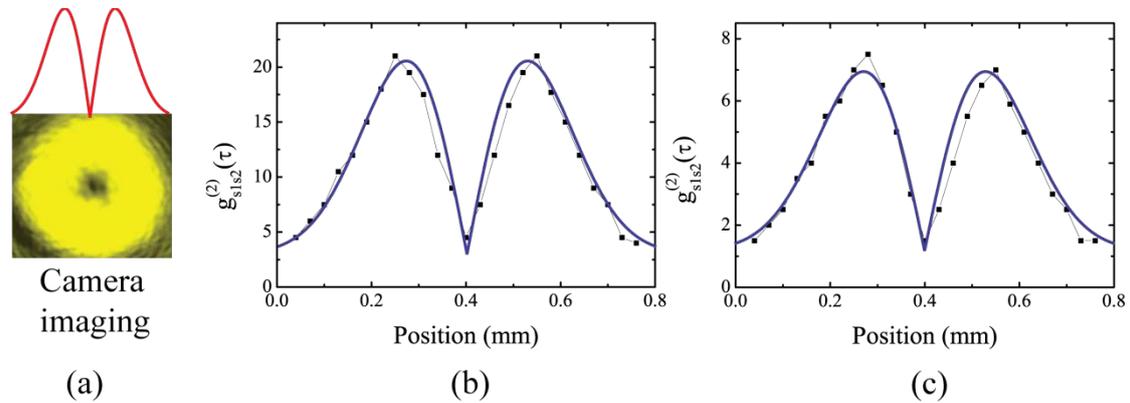

(a)                           (b)                          (c)

Fig. 2 (Color online) (a) Image of a laser beam passing through the spiral phase plate. (b) the cross-correlation between the input signal and the trigger photons, obtained by scanning the transverse position of the input signal. (c) the cross-correlation function between the retrieved signal and the trigger photons. The solid lines were the theoretical fitted curves.

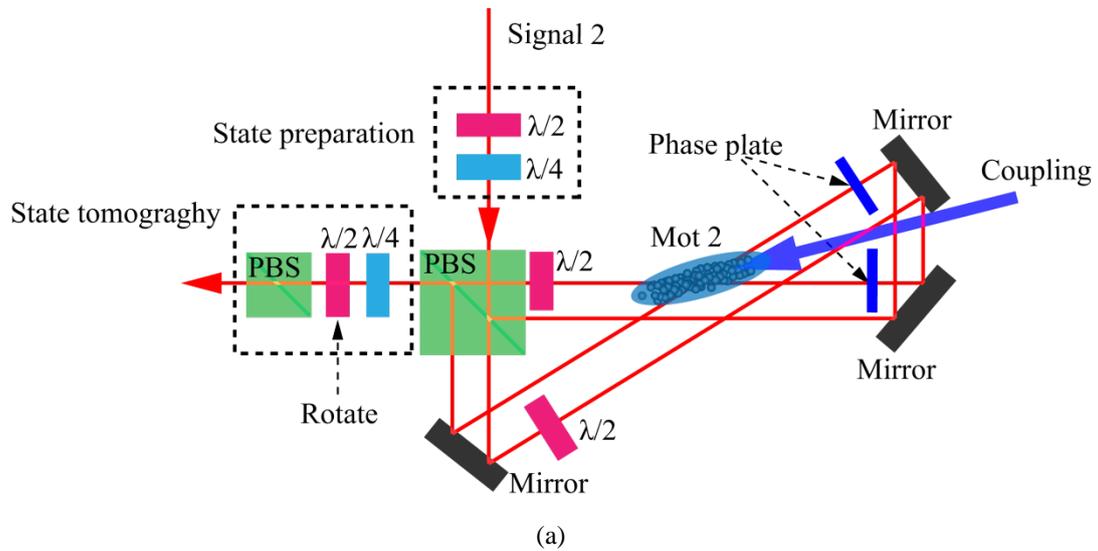

(a)

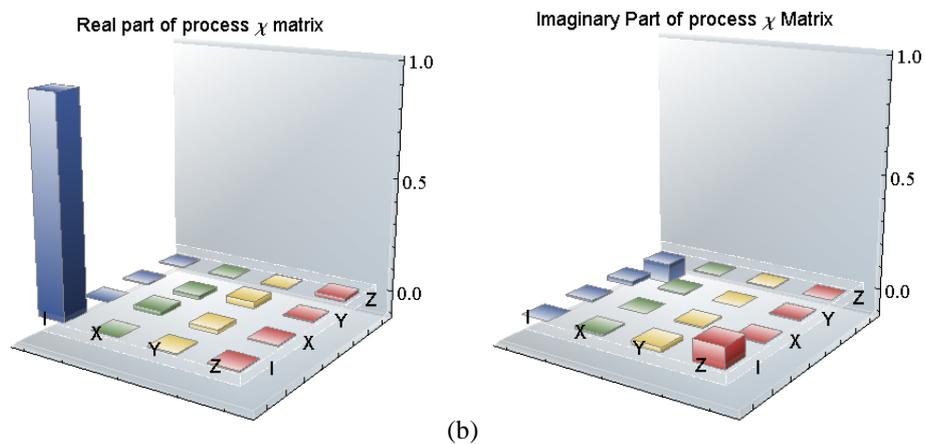

(b)

Fig. 3. (a) The simplified experimental setup for proving the coherence of single photon. (b) The calculated real part and the imaginary part of the storage process matrix according to experimental data.

For the useage of our scheme in practice, we have to prove that the property of the coherence of the photon is preserved through the storage. Here we made two experiments to demonstrate it. Firstly, we experimentally checked whether we could store an arbitrary polarization state of the input signal photon. In order for that, we performed the quantum storage process tomography [32] for this storage process, which was realized by the aid of a Sagnac interferometer shown in Fig. 3(a). In this process, two phase plates inserted in the interferometer were removed. The whole setup consisted of three parts: state preparation part for preparaing an arbitrary polarization state; Sagnac interferometer for storing the polarization state and the state tomography part for state analysis. In this experiment, the photon did not carry a spatial structure. Using such a configuration, two orthogonal polarization parts of an input state were stored in the atomic ensemble along forward and backward directions respectively. The experimental detail was shown in method section. Fig. 3(b) was the process matrix $\chi$ constructed according to experimental data. The calculated the fidelities of the storage process were 0.94, 0.96, 0.98, and 0.96 for four different input polarization states of H, V, R, and D, respectively, where H and V stand for horizontal and vertical polarizations, $R = \frac{H - iV}{\sqrt{2}}$, right-circular polarization, and $D = \frac{H + V}{\sqrt{2}}$, diagonal polarization. In this process, the storage time was programmed to be 100 ns. The experimental results clearly demonstrated the preservation of the coherence of the photon during the storage process. The use of the Sagnac interferometer in our experiment avoided the phase flucturation between two orthogonal polarization parts of the signal. The detail about the experiment was shown in method section too

In the second experiment, we experimentally checked whether we could store an OAM superposition state of the input signal photon. In order for that, we inserted the two phase plate (VPP-2, RPC Photonics) into the two optical routes (see Fig. 3(a)), the prepared state of input signal photon for the storage was

$$|\psi\rangle = \frac{1}{\sqrt{2}}[|H\rangle|l\rangle + |V\rangle|-l\rangle], \tag{1}$$

In our experiment, $l=2$. The outptut state became to be $|\psi_1\rangle=\frac{1}{\sqrt{2}}(|L\rangle|l\rangle+i|R\rangle|-l\rangle)$ when the retrieved signal passed through a quarter-wave plate with 45° rotation. Then we used a half-wave plate (shown in Fig. 3(a)) to rotate the polarization of the photon, the output state was:

$$|\psi_2\rangle = \hat{U}_{HWP}(\theta)|\psi_1\rangle \qquad (2)$$

where $\hat{U}_{HWP}(\theta)$ was the transformational matrix of a half-wave plate. $\theta$ was the angle of the fast axis with respect to the vertical axis. The output state became to be:

$$|\psi_3\rangle = \frac{1}{\sqrt{2}}(e^{-i2\theta}|-l\rangle + ie^{i2\theta}|l\rangle) \quad . \qquad (3)$$

This result told us that there was an interference between two terms in Eq. 3, which gave a characteristic pattern comprising 4 spots. Shifting the relative phase of the two terms caused the spot pattern to rotate. In our experiment, shifting phase was done by rotating the half-wave plate, i.e. changing $\theta$ continuously. In the experiment, we firstly used a weak coherent light (~$10^4$ photons per pulse) as the input signal. We set the angle of the half-wave plate to be $\theta=22.5°, 67.5°, 112.5°$ and $157.5°$ respectively, and used a camera (CCD, 1024×10, iStar 334T series, Andor) to monitor the spatial shape of the light. The results were shown in Figure. 4(a). The theoretical simulations and the experimental data agreed very well. Moreover, we coupled the photons in one spot into a fiber and detected them by a photon detector (Avalanche diode, PerkinElmer SPCM-AQR-15-FC). From Eq. 3 we knew that $|l\rangle$ and $|-l\rangle$ had an additional phase $2\theta+\frac{\pi}{2}$ and $-2\theta$ respectively. With regard to the interference pattern in one spot, the detected intensity will change with a function of $sin(4\theta)$. Therefore we should obtain a sine-shaped interference pattern when we changed the angle of the half-wave plate. In this experiment, we used the true single photon (signal 2 photon, generated through SFWM in MOT 1) as the input signal. The intergrated coincidence counts per second in 50 ns coincidence window with the substracted background noise showed the clear interference pattern against the angle of the half-wave plate as in Fig. 4(b). The visibility of this interference was 0.74±0.1, the error bar was the statistics in our experiment, which was mainly from the noise of coupling laser and the multi-photon events. The experimental results were in well agreement with our theoreticl predictions. In this process, the storage time was programed to be 100 ns.

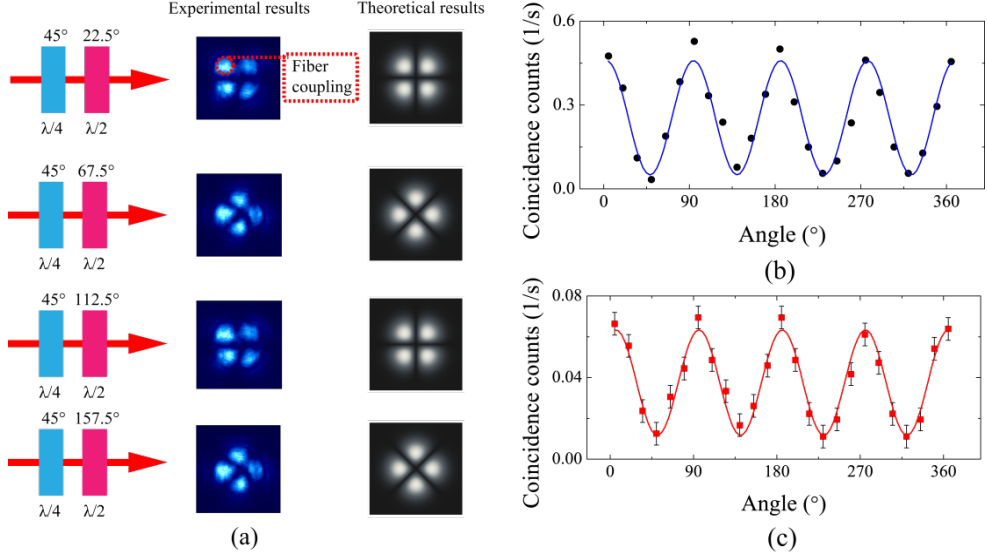

Fig 4 (a) the rotated interference pattern with different angle of the half-wave plate. (b) the interference pattern of input signal photon against the angle of half-wave plate. (c) the interference of retrieved signal against the angle of half-wave plate. The blue and red lines were the fitted curves with $sin(4\theta)$ function.

The previous results obtained clearly demonstrated that the coherence between two different OAM states was preserved during the storage. Combined with the results shown in Fig. 3, we can conclude that the property of the coherence of the photon is preserved through the storage.

In our work, we gave an experimental verification of the ability of storing a true single photon carrying a spatial structure in a cold atomic ensemble for the first time. Our work made an important step toward realizing a high-dimensional quantum memory using an atomic ensemble. Along with the recent important progresses made in the areas of infrared-to-visible wavelength conversion [33] and long-distance fiber transmission of a photon encoded in a high-dimensional state [34], or with the significant progresses made in the area of the quantum key distribution between the ground and air [35, 36]. Our results may lay the basis for establishing a high-dimensional quantum network in the future.

**Method sections**
**Experimental preparation of a true single photon**

See Fig. 1(a). A cigar-shaped atomic cloud of $Rb^{85}$ atoms obtained in MOT 1 was used to prepare a real single photon via the SFWM in a double-lambda configuration, where the states we used were|1>,

|2>, |3> and |4>, corresponding to the energy levels of $5S_{1/2}$ (F=3), $5S_{1/2}$ (F=2), $5P_{3/2}$ (F'=1) and $5P_{1/2}$ (F'=3) respectively. The pump 1 was from an external-cavity diode laser (DL100, Toptica), had the wavelength of 780 nm, and was set to be red-detuned 50 MHz to the atomic transition of $5S_{1/2}$ (F=3) -> $5P_{3/2}$ (F'=1). The pump 2, from another external-cavity diode laser (DL100, Toptica), having the wavelength of 795 nm, was set to be resonant on the atomic transition of $5S_{1/2}$ (F=2) -> $5P_{1/2}$ (F'=3). The pump 1 and pump 2 had opposite linear polarizations, the signal 1 and signal 2 obtained through SFWM also had opposite linear polarizations. The powers of pump 1 and pump 2 were 54 μW and 0.65 mW respectively. The optical depth (OD) of MOT 1 was about 8. Under the conditions of phase matching and the conversation of energy, the generated signal 1 and signal 2 photons were non-classical correlated in time domain. In the experiment, signal 1 and signal 2 photons were coupled into two single mode fibers with a coupling efficiency of 90%. Before that, an interference filter with a transmission coefficient of 50% was used to reduce the noise in the route that signal 1 transmitted along. Besides, two homemade Fabry-Perot Etalons (FP) were inserted into the optical routes that signal 1 and signal 2 photons transmitted along to further reduce the noise. Each FP hada transmission efficiency of 83% and a 500-MHz bandwidth. After that, signal 1 photon was detected by a single photon detectors (Avalanche diode, PerkinElmer SPCM-AQR-15-FC with 50% efficiency.) and the signal 2 photon was used for the storage in the follow experiments. The retrieved signal 2 photon after been stored for a given time was detected by another single photon detector (SPCM-AQR-15-FC). The outputs of detectors were connected to a time-to-digital converter (Fast Comtec. P7888) with 1 ns bin width to measure the cross-correlation function $g_{s1,s2}(\tau)$.

Usually classical lights satisfied the following equation [26]:

$$R = \frac{[g_{s1,s2}(\tau)]^2}{g_{s1,s1}(0)g_{s2,s2}(0)} \leq 1, \qquad (4)$$

where, $g_{s1,s2}(\tau)$, $g_{s1,s1}(0)$ and $g_{s2,s2}(0)$ were the normalized second-order cross-correlation and auto-correlation of the photons respectively. The normalized $g_{s1,s2}(\tau)$ can be obtained by normalizing the true two-photon coincidence counts to the accidental two-photon coincidence counts $g_{s1,s2}(\infty)$. $\tau = t_{s1} - t_{s2}$ is the relative time delay between paired photons. For instance, the maximum $g_{s1,s2}(\tau)$ we obtained in the experiment was $g_{s1,s2}(\tau) = 200 \pm 4$ at $\tau = 19$ ns, thus the corresponding Cauchy-Schwarz inequality factor $R = 10000 \pm 400$ was much larger than 1 using the fact that $g_{s1,s1}(0) = g_{s2,s2}(0) \approx 2$ (signal photons 1 and 2 exhibited the thermal light photon statistics), then the Cauchy-Schwarz inequality was strongly violated, which clearly demonstrated the non-classical correlation between photons. The full width at half maximum of the cross-correlation function was 32.5 ns, so we estimated that the frequency full band width at half maxim of the photon was ~30 MHz. The photon bandwidth could be tuned for example by changing the Rabi frequency of the pump beam [37].

For an ideally prepared single photon state, $\alpha \rightarrow 0$, for a classical field, $\alpha \geq 1$, based on the Cauchy-Schwarz inequality [27, 28]. Here the anti-correlation parameter $\alpha = \frac{P_1 P_{123}}{P_{12} P_{13}}$, was given by the ratio of various photoelectric detection probabilities which were measured by the set of detectors $D_1$, $D_2$ and $D_3$. $P_1$ was the trigger photon counts, $P_{12}$ and $P_{13}$ were the twofold coincidence counts between the trigger and the two separated signal photon (the signal was divided into two equal parts by a beam splitter, and detected by two single photons detectors $D_2$ and $D_3$ respectively. These were not shown in Fig. 1.) and $P_{123}$ was the threefold coincidence counts between three detectors $D_1$, $D_2$ and $D_3$.

**Experimental storage of a true single photon**

The experimental storage of a single photon through EIT was implemented in another MOT 2, shown in Fig. 1(b). The OD of the atomic vapor in MOT 2 was about 10. The bandwidth of storage was about ~20 MHz. The coupling laser, which was from the same laser with the pump 1 beam, coupled the atomic transition of $5S_{1/2}(F=2) \rightarrow 5P_{1/2}(F'=3)$. The signal was the photon needed to be stored, and the trigger was detected by a detector (detector 1). The output electrical signal from detector 1 was input into an arbitrary function generator (AFG 3252) to trigger an acousto-optic modulator (AOM) to switch on/off the coupling laser with a rising/falling edge of 15 ns. The signal firstly transmitted through a 200-metre long single mode fiber, then was focused into the atomic cloud achieved in MOT 2 using a lens with a focal length of 500 mm. The coupling beam was focused by a lens with a focus length of 700 mm and it can cover the probe beam completely. The Rabi frequency of coupling laser was $4\Gamma$, where $\Gamma$ was decay rate of level $|4\rangle$. The angle between the coupling and probe was 3°. The non-collinear configuration used in the experiment could significantly reduce the noise from the coupling beam scattering. The retrieved signal was coupled into a single mode fiber and was detected by another detector. We carried out our experiment with the repeatability of 100 Hz. The atomic trapping time was 8 ms, the experimental window was 1.5 ms, another 0.5 ms was used to prepare the initial state.

**Experimental quantum process tomography of a true single photon**

The experimental conditions of the quantum process tomography [32] of arbitrary polarizations were similar to the previous experiment, beside the follows: about photon pair source based on SFWM, the powers of pump 1 and pump 2 were 120 μW and 1 mW respectively. The pump 1 was set to be red-detuned 170 MHz to the atomic transition of $5S_{1/2}(F=3) \rightarrow 5P_{3/2}(F'=1)$. In the experiment, the efficiency of coupling trigger and signal photons into two single mode fibers was of 85%. The previous interference filter was replaced of two new interference filters with transmission efficiency of >95% at 780 nm. In order to reduce the noise from the pump 2 and coupling laser beams, the previous FP (inserted in signal optical route) was replaced by two new Fabry-Perot Etalons (FP) (one FP transmission efficiency was 83%, another was 90%). The full band width at half maxim of the photon was ~13MHz. About storage process in the second MOT, the signal photon was focused into the second atomic cloud using a lens with a focal length of 1.9 m. The Rabi frequency of coupling laser was 5Γ. The coupling beam was an elliptical laser beam with the size of 1×3.5 mm in the atomic ensemble and it can cover the probe beam completely. The retrieved signal photon was coupled by a single mode optical fiber with efficiency of 50%. In order to reduce this noise, we inserted two FPs to reduce the noise from the coupling beam with an attenuation rate of ~50000.

**Acknowledgements**

We thank Dr. Quentin Glorieux very much for carefully reading this manuscript and giving many useful comments and advices. We also thank Dr. Jiang-Feng Du and Mr. Peng-Fei Wang for kindly loan us spiral phase plate. We thank for discussions with Mr. Xian-min Jin and Mr. Zong-quan Zhou, and Dr. Guo-yong Xiang. This work was supported by the National Natural Science Foundation of China (Grant Nos. 11174271, 61275115, 10874171), the National Fundamental Research Program of China (Grant No. 2011CB00200), the Youth Innovation Fund from USTC (Grant No.ZC 9850320804), and the Innovation Fund from CAS, Program for NCET.